\documentstyle[aps,prl,preprint]{revtex}
\tightenlines
\newcommand{\etal}{{\em et al. }}                

\newcommand{\eqref}[1]{(\ref{#1})}
\newcommand{\dn}[2]{\;d^{#1}{#2}\;}
\newcommand{\twospectrum}[2]{E_1 E_2 \frac{dN^{#1}_{#2}}{d\vec{p}_1d\vec{p}_2}}
\newcommand{\twospectrumalt}[2]{dN^{#1}_{#2}/d\vec{p}_1d\vec{p}_2}
\newcommand{\twotrue}{\twospectrum{\rm true}{}}
\newcommand{\twomixed}{\twospectrum{\rm mixed}{}}
\newcommand{\twotruealt}{\twospectrumalt{\rm true}{}}
\newcommand{\twomixedalt}{\twospectrumalt{\rm mixed}{}}
\begin{document}
\title{
  \begin{flushright}{\rm DOE/ER/40561-88-INT00}\\[9mm]\end{flushright}
  Is it possible to reconstruct the freeze-out duration of heavy-ion 
collisions using tomography?}
\author{David A. Brown}
\address
{Institute for Nuclear Theory, 
University of Washington, Box 351550, Seattle, WA 98195-1550}
\date{\today}
\maketitle
%
\begin{abstract}
We investigate what conditions allow us to extract the relative distribution of 
freeze-out space and time points in an arbitrary reference frame using 
tomography and source imaging.  
The source function may be extracted from the two-particle 
correlation function measured in heavy-ion collisions using imaging 
techniques.  This imaged source function is related 
to the relative distribution of freeze-out space and time points through 
a generalization of the Radon transform found in tomography.
Using tomography, the imaged source function may be converted 
into the relative freeze-out distribution in the frame of interest.  
We describe how the tomography may be performed in practice.

\end{abstract}
%

We intend to create the quark-gluon plasma in central nuclear collisions
at the Relativistic Heavy-Ion Collider (RHIC).  The quark-gluon plasma is 
predicted to have a long lifetime~\cite{Rischke:1996cm,Bass:1999vz} which 
should lead to large time separations between the emission (or freeze-out)
times of pairs 
of like particles.  As two-pion correlations are sensitive to this time
separation, they should be a useful tool for studying the 
plasma~\cite{Bass:1999vz,lifetimeHBT}.
The signal of a long lifetime would be   
anomalously large correlation radii, particularly in the longitudinal and 
outward directions (in a analysis done in the Bertsch-Pratt parameterization
\cite{BPcoords}).
The lifetime information is encoded in these directions in a non-trivial 
manner which is currently accessed only using model parameterizations
of two-pion correlation functions~\cite{Bass:1999vz}.
A long lifetime should also effect other like-pair 
correlations, such as two-proton or two-kaon 
correlations~\cite{koonin_77,gelbke_90,bauer_92,pratt_98}.  
However as in the case of protons, final state interactions often obscure
this information.

Recently it was shown that one can perform model-independent 
extractions of the {\em entire} source function $S(\vec{r})$ from two-particle 
correlations, not just its radii, using imaging techniques
\cite{HBT:bro97,HBT:bro98,HBT:bro98a}. 
Furthermore, one can do this even with relatively complicated
final-state interactions (e.g. protons) and without making any {\em a priori} 
assumptions about the source geometry or lifetime, etc.
First results from the application of imaging to the
proton, pion, kaon, and IMF correlation data can be found in 
Refs.~\cite{HBT:bro97,HBT:bro98,HBT:bro98a,pp:e895}. 
These results are both intriguing and nearly as hard to interpret as 
the original correlation functions:  the reconstructed
sources are found in the pair Center-of-Mass (CM) frame and the time 
dependence is folded into the sources in a non-trivial way.

These two ambiguities are intimately tied together: the time direction is
folded into the source function by a line integral in the 
direction of the boost from the system (or lab) frame to the CM frame. 
As we show in Eq.~\eqref{sourcetocalStilde}, this line integral is a 
generalized Radon transform of the relative distribution of space-time 
emission points.  If we construct the correlation function  
from pairs with a fixed rapidity and transverse momentum, then we 
know the boost from the system frame to the pair CM frame.  
Therefore, it should be possible to use tomography to reconstruct the relative 
distribution of space-time emission points from the imaged sources and we give 
an explicit inversion formula in Eq.~\eqref{inverse}.  There is one caveat: 
for the reconstruction to be unique, we must require 
that the probability for emitting a pair with a certain relative separation be 
independent of the total pair momentum.  In what follows, we will 
refer to this condition as the momentum averaging 
approximation.  The experimental signal for this condition to hold would be 
a sideward radius parameter that does not depend on  the total pair momentum
and this condition seems to be fulfilled at the CERN-SPS~\cite{NA49stuff}.
This analysis may be useful in the correlation function analysis of 
other systems, such as fragmenting hadronic strings or 
relativistic atomic collisions.

The outline of this letter is as follows.  First, we detail the 
imaging of the source function as a function of pair momentum 
and show that the imaged source function is a generalized
Radon transform of the full two-particle source.  
Second, we study the reconstruction of the relative 
space-time distribution of emission points from the imaged source.
Finally, we outline how to perform this reconstruction in a practical
manner.  

We begin with a discussion of the correlation function and its relation 
to the source function.  For concreteness, we consider like-pairs from central
relativistic nucleus-nucleus reactions.  We are eventually interested in 
quantities in the lab frame, namely the center-of-mass frame of the colliding
nuclear system.  The pairs we consider can be protons, pions, etc., as the 
formalism works for any like-pairs.
The like-pair correlation function in an 
arbitrary frame is the following ratio of invariant spectra:
\begin{equation}
  C(P,q)=\displaystyle\twotrue\left/\displaystyle\twomixed\right..
  \label{anycorr}
\end{equation}
Here, $\twotruealt$ is the two-particle spectrum of pairs from the same event
(averaged over many events) and $\twomixedalt$ is the two-particle spectrum 
constructed from pairs in different events.  In this expression, $\vec{p}_1$ 
and $\vec{p}_2$ are the three-momentum of each of the particles 
and $E_1$ and $E_2$ are the on-shell energies.  For simplicity, we work in 
side--out--long coordinates (a.k.a. Bertsch-Pratt \cite{BPcoords}) 
coordinates where
the longitudinal axis is along the beam line, the outward axis is along
the component of $\vec{P}$ perpendicular to the longitudinal direction,
and the sideward axis is perpendicular to the other two directions.  In these
coordinates, the total 
four-momentum of the pair is $P=p_1+p_2=(E,P_L,P_O,0)$ and the relative 
four-momentum is $q=\frac{1}{2}(p_1-p_2)=(q_0,q_L,q_O,q_S)$.

In the Koonin-Pratt formalism, the correlation in Eq.~\eqref{anycorr} 
is related to the normalized single-particle sources through
\cite{koonin_77,HBT:bro97,HBT:bro98,HBT:pra90}:
\begin{equation}
  C(P,q)=\int \dn{4}{r} |\Phi^{(-){\rm rel}}(r,q)|^2\int\dn{4}{R} 
  \tilde{D} (R+r/2,P/2+q)\tilde{D} (R-r/2,P/2-q)
  \label{KPeqn1}
\end{equation}
Here $\Phi^{(-){\rm rel}}(r,q)$ is the pair relative wavefunction of the 
emitted pair and it includes all final-state interactions between the pair as
well as (anti-)symmetrization due to statistics.  Also 
$\tilde{D}(r,p)$ is the normalized single-particle source and it denotes the
probability of emitting an on-shell particle with four-momentum $p$ at 
position $\vec{r}$ at time $t$.  
In terms of the emission rate, $\tilde{D}(r,p)$ is:
\begin{equation}
   \tilde{D}(r,p)=\displaystyle\frac{E\dn{7}{N}}
   {\displaystyle\dn{3}{p}\dn{4}{r}}\left/
   \displaystyle \frac{\displaystyle E\dn{3}{N}}{\displaystyle\dn{3}{p}}\right.
\end{equation}
where the emission rate is the number of particles frozen out per
unit time, per unit volume, per unit volume of invariant momentum.
Note that the normalized single-particle source transforms as a 
four-scalar.  

We may measure the correlation function and write Eq.~\eqref{KPeqn1} in 
whatever frame we wish.  In the pair center-of-mass (CM) frame 
Eq.~\eqref{KPeqn1} simplifies considerably:
\begin{equation}
  C_{\vec{P}}(\vec{q'})=\int \dn{4}{r'}|\Phi^{(-){\rm rel}}
      (\vec{r'},\vec{q'})|^2\int\dn{4}{R'} 
      \tilde{D}_{\vec{P}} (R'+r'/2,\vec{q'})\tilde{D}_{\vec{P}} 
      (R'-r'/2,-\vec{q'}).
  \label{KPeqn2}
\end{equation}
The subscript $\vec{P}$ refers to the momentum of the boost from some other 
frame (e.g. the lab frame) to the pair CM frame and the primed coordinates are
pair CM coordinates.  Note that in the pair CM frame $\vec{P'}=0$ and 
$E'=2m$.  The reader should also note that {\em each pair has a 
different $\vec{P}$ so it is boosted to a 
different pair CM frame}. Furthermore, because of our choice of coordinates,
this boost is confined to the longitudinal and outward directions.
In order to remove the $t'$ dependence of the wavefunction from 
Eq.~\eqref{KPeqn2}, we have taken advantage of the facts that 
the particles are on-shell, giving $q_0=\vec{\beta}\cdot\vec{q}$,  
and that our wavefunction is a function of the Lorentz scalar 
$q\cdot r=(t \vec{\beta}-\vec{r}\,)\cdot\vec{q}=-\vec{r'}\cdot\vec{q'}$.

We now make two crucial assumptions: the {\em smoothness approximation} and a
{\em momentum averaging approximation}.  For the smoothness assumption, we 
assume that the normalized single-particle 
sources have only a weak dependence on $\vec{q'}$.  This is justified because
the single-particle source varies weakly in $\vec{r'}$ as a function of 
$\vec{q'}$ while the wavefunction oscillates rapidly in $\vec{r'}$ for large 
$\vec{q'}$~\cite{HBT:pra90,Heinz_review}.  Thus, at low $\vec{q'}$ 
the single-particle source is approximately independent of 
$\vec{q'}$ while at large $\vec{q'}$ the integral of the source function 
and the wavefunction average to zero.  This approximation allows us to 
define the source function \cite{HBT:bro97,HBT:bro98}:
\begin{equation}
   S_{\vec{P}}(\vec{r'})=\int \dn{}{t'}
   \int \dn{4}{R'}\tilde{D}_{\vec{P}} (R'+r'/2,\vec{q'}\approx 0)
                  \tilde{D}_{\vec{P}} (R'-r'/2,\vec{q'}\approx 0).
   \label{defofsource}
\end{equation}
$S_{\vec{P}}(\vec{r'})$ is the probability of creating a pair a distance 
$\vec{r'}$ apart in the pair CM frame.
In terms of the source function, Eq.~\eqref{KPeqn2} becomes
\begin{equation}
     C_{\vec{P}}(\vec{q'})=\int \dn{3}{r'}
	|\Phi^{(-){\rm rel}}(\vec{r'},\vec{q'})|^2 S_{\vec{P}}(\vec{r'}).
     \label{KPeqn3}
\end{equation}
Since the source function in Eq.~\eqref{KPeqn3} is independent of the relative
pair momentum, we can uniquely invert $C_{\vec{P}}(\vec{q'})$ 
to obtain $S_{\vec{P}}(\vec{r'})$ using the 
imaging techniques of Refs.~\cite{HBT:bro97,HBT:bro98}.  Therefore, we 
can dispense with the correlation function and work directly with the imaged 
source function.

For reasons that will become clear momentarily, let us define the 
two-particle source as the probability of creating a pair of 
particles space-time distance of $r=(t,\vec{r})$ apart in the lab frame.  
We may write the the two-particle source ${\cal S}(r,P)$ 
in terms of the normalized single-particle sources as
\begin{equation}\begin{array}{rl}
   {\cal S}(r,P)&=\displaystyle\int \dn{4}{R}
  \tilde{D}(R+r/2,P/2)\tilde{D}(R-r/2,P/2)\\
  &=\displaystyle
  \int \dn{4}{R'}\tilde{D}_{\vec{P}} (R'+r'/2,\vec{q'} \approx 0)
  \tilde{D}_{\vec{P}} (R'-r'/2,\vec{q'}\approx 0).
\end{array}\label{defofcalS}\end{equation}
The second line in this equation is just the first line rewritten in the 
pair CM frame.  The space-time displacement $r'$ in the second line should be
understood as a function of the boost velocity and space-time displacement
$r$ in the original frame.  Here we have dropped the $q$ dependence in both 
the lab frame (as in the lab frame $P\gg q$) and in the pair CM frame 
(in the smoothing approximation).   Comparing the 
definition of the source function~\eqref{KPeqn2} with the second line
of Eq.~\eqref{defofcalS}, we see
\begin{equation}
   S_{\vec{P}}(\vec{r'})=\int \dn{}{t'}{\cal S}(r,P).
   \label{sourcetocalS}
\end{equation}
Thus, the imaged source is a line integral of the two-particle source along
some boosted direction given by the pair total momentum, $\vec{P}$.

In the momentum averaging approximation, we
replace ${\cal S}(r,P)$ in Eq.~\eqref{sourcetocalS} 
with $\tilde{\cal S}(r)$.  Here the $\tilde{\cal S}(r)$ has the 
interpretation as the relative distribution of space-time emission points,  
i.e. it gives the probability density for emitting any pair from 
space-time points separated by a displacement $r$ in the lab frame.
In terms of this distribution, we write Eq.~\eqref{sourcetocalS} as
\begin{equation}
   S_{\vec{P}}(\vec{r'})=\int \dn{}{t'}\tilde{\cal S}(r).
   \label{sourcetocalStilde}
   \label{marginaldist}
\end{equation}
Here all of the $\vec{P}$ dependence of the source function is 
attributed to the boost from the lab frame to the pair CM frame.  This 
transform may be regarded as a generalized Radon transform as it involves
boosts instead of rotations.  In the conventional Radon 
transform~\cite{cantget,natterer,tarantola}, one integrates
a function $f(x,y)$ along a direction rotated at an angle $\theta$, e.g.
\begin{equation}
   f_\theta(x')=\int \dn{}{y'} f(x,y)
\end{equation}
where $y'=x\sin{\theta}+y\cos{\theta}$ and $x'=x\cos{\theta}-y\sin{\theta}$.
In Eq.~\eqref{sourcetocalStilde}, we are integrating along a time direction
in a boosted frame.  Thus, $t'$ in Eq.~\eqref{sourcetocalStilde} plays the role 
of $x$ in the conventional Radon transform and $\vec{P}$ (or equivalently the 
boost velocity) plays the role of the rotation angle $\theta$.

The momentum averaging approximation is justified for the following reasons.
First, we could just {\em define} the relative distribution of emission 
points through Eq.~\eqref{marginaldist}.  In this case, we will be folding 
any $P$ dependence (such as from collective motion) into the time direction.  
Alternatively,  we may think of $\tilde{\cal S}(r)$ as ${\cal S}(r,P)$, 
but averaged over $P$.   This is the viewpoint we adopt in this letter.
Finally, in some systems or situations the two particle source only has a weak
dependence on $P$ and we may neglect this dependence.  
In a heavy-ion reaction, we can check this approximation by 
measuring the (e.g. pion) correlation function as a function of total pair
momentum.  Then we must examine the dependence of the correlation in the 
sideward and longitudinal/outward directions.
If the dependence of the correlation in the sideward direction is 
relatively flat as a function of the total pair momentum, the two particle 
source does not have a $P$ dependence.  We will illustrate this below 
in a simple Gaussian model.  We comment that the sideward pion
radius parameter measured by the CERN-SPS experiment NA49~\cite{NA49stuff} 
is relatively flat as a function of pair rapidity and transverse momentum so 
the momentum averaging approximation seems justified.  
Strictly speaking, comparing the outward and sideward direction tells us that
there is no $P$ dependence of ${\cal S}(r,P)$ 
in the sideward/outward directions, but tells us
nothing about the $P$ dependence in the longitudinal direction.  
We postpone the detailed investigation of this 
approximation (especially in the presence of collective motion)
for a future article~\cite{tomogtodo}.

Let us now illustrate how the transform in Eq.~\eqref{sourcetocalStilde} works 
on a model two-particle source with time evolution, but no dependence on the 
total pair momentum.
Assume that we have spherical stationary two-particle source with radius $R$ and
time duration $\tau$ in the lab frame:
\begin{equation}
   {\cal S}(t,\vec{r}) = 
   {\cal S}_0\exp\left(-\frac{t^2}{2\tau^2}-\frac{\vec{r}^2}{2R^2}\right).
\end{equation}
Since ${\cal S}(t,\vec{r})$ denotes the probability density for producing the 
pair with a space-time separation $(t,\vec{r})$, the normalization 
constant is ${\cal S}_0=(4\pi^2 \tau R^3)^{-1}$.  The boost from the 
lab to the pair CM frame is characterized by the boost velocity 
$\vec{\beta}$, which may be written in terms of the total pair momentum giving 
$\vec{\beta}=\vec{P}/E\equiv (\beta_L,\beta_O,0)$.  
The coordinates transform from the pair CM frame to the system frame via
$\vec{r}=\vec{r'}+\frac{\gamma-1}{\beta^2}(\vec{\beta}\cdot\vec{r'})\vec{\beta}
+\gamma\vec{\beta}t'$ and $t=\gamma(t'+\vec{\beta}\cdot\vec{r'})$ with
$\gamma=1/\sqrt{1-\vec{\beta}^2}$.
Integrating over the CM time to find the imaged source function,
we find:
\begin{equation}
	S_{\vec{\beta}} (\vec{r'})={\cal S}_0 \frac{\sqrt{2\pi}}{\gamma}
	\sqrt{\frac{\tau^2 R^2}{R^2+\beta^2\tau^2}}\exp\left(-\frac{1}{2R^2}
	\left(\vec{r'}^2-(\vec{\beta}\cdot\vec{r'})^2
	\frac{R^2+\tau^2}{R^2+\beta^2\tau^2}\right)\right).
\end{equation}
This source function is also Gaussian and we immediately see how the 
radius parameters are modified by the boost. First,  since $\vec{\beta}$ has
no component in the sideward direction, the Gaussian radius in the sideward 
direction is independent of $\vec{\beta}$ and is equal to the radius of the 
original two-particle source, $R_S=R$.  We remind the reader that a flat 
sideward radius parameter is a signal of the validity of the momentum 
averaging approximation.
Second, both the outward and longitudinal radius parameters become
$R_{O/L}=R/\sqrt{1-\beta^2_{O/L}(R^2+\tau^2)/(R^2+\beta^2\tau^2)}$.
So for large emission durations, we find anomalously large radius parameters 
in the outward and longitudinal directions.  Finally, we find a long/out cross 
term $R^2_{OL}=-R^2/\beta_L\beta_O\sqrt{(R^2+\tau^2)/(R^2+\beta^2\tau^2)}$.

Now we can invert Eq.~\eqref{marginaldist} and obtain the inverse 
in the Bertsch-Pratt coordinates:
\begin{equation}\begin{array}{lr}
   \lefteqn{\tilde{\cal S}(t,r_L,r_O,r_S)=}&\\
   &\displaystyle\frac{1}{(2\pi)^3}\int\dn{}{r'_L}\dn{}{r'_O}
   \int\dn{}{\beta_L}\dn{}{\beta_O}
   g\left(\hat{\beta}\cdot(\vec{r'}-\gamma(\vec{r}-t\vec{\beta}))\right)
   \frac{\gamma^3}{\beta}S_{\vec{\beta}}(r_L',r_O',r_S').
\end{array}\label{inverse}\end{equation}
Here $g(z)=\int^{\infty}_{-\infty}\dn{}{\mu}\mu^2e^{-i\mu z}$
is a universal filter function.  Due to our choice of 
coordinates, $r_S=r_S'$.  The derivation of the inverse in~\eqref{inverse} 
is straightforward following the appendix 
of Ref.~\cite{janicke95}, so we postpone the detailed derivation for 
a future article~\cite{tomogtodo}.

Let us now comment on inverting Eq.~\eqref{sourcetocalStilde} 
in practice.  We might just plug the imaged source into
Eq.~\eqref{inverse} and integrate to find the
relative distribution of space-time points.  This method is called the
filtered back-projection algorithm and  
it is a poor way to perform the inversion in practice 
because of the singular behavior of the filter function $g$.
A regularization of the filter function is possible, but introduces 
numerical inaccuracy~\cite{natterer}.  Furthermore, the filtered 
back-projection algorithm is computationally intensive owing to the four 
integrals in Eq.~\eqref{inverse}.  

A better approach is to use
Algebraic Reconstructive Tomography (ART)~\cite{natterer,tarantola}.  
In ART, one expands both the imaged source 
function and the relative distribution of space-time emission points in
Eq.~\eqref{sourcetocalStilde} in some function basis. For the sake 
of illustration we will use Basis splines~\cite{deboor} for this basis.  
In this basis,
$S_{\vec{P}}(\vec{r'})=\sum_{ij} S_{ij} B_i(\vec{P}) B_j(\vec{r'})$ and
 $\tilde{\cal S}(r)=\sum_k \tilde{\cal S}_k B_k(r)$ where 
$i, j,$ and $k$ are indices in the directions of
$\vec{P}, \vec{r'},$ and $r$ respectively.  
Eq.~\eqref{sourcetocalStilde} may then be recast as
\begin{equation}
   \sum_{ij} S_{ij} B_i(\vec{P}) B_j(\vec{r'})=
   \int \dn{}{t'} \sum_k \tilde{\cal S}_k B_k(r).
\end{equation}
Since the Basis splines are orthogonal, one can write this as a matrix equation:
\begin{equation}
	S_{ij}=\sum_k \tilde{\cal S}_k \int \dn{3}{P} \dn{3}{r'} 
	B_i(\vec{P}) B_j(\vec{r'}) \int \dn{}{t'} B_k(r)
\end{equation}
Only the $t'$ integral needs to be done numerically as the other six integrals 
may be done analytically.  Thus, ART reduces to inverting a matrix equation 
for the coefficients of the Basis spline expansion.
We will explore the viability of the this approach in a future 
article~\cite{tomogtodo}.


In conclusion, we should be able to reconstruct the relative distribution
of emission space and time points from the correlation functions 
measured in heavy-ion collisions.  Our approach is model-independent and works 
for any like-pair correlations as it is implemented at the level of source 
functions.  The momentum averaging approximation involved in the reconstruction
seems to be justified in some experiments and its validity can be checked 
experimentally.  Thus, this reconstruction provides us with a way to directly 
measure the freeze-out duration of heavy-ion collisions.
%
\section*{Acknowledgements}
\indent We wholeheartedly thank Drs. G.~Bertsch, J.~Cramer, P.~Danielewicz 
and T.~Papenbrock
for their thorough reading of the manuscript.  We also 
acknowledge stimulating discussions with H. Vija and Drs. J.~Cramer, 
G.~Bertsch, P.~Danielewicz, S.~Panitkin, T.~Papenbrock,
A.~Parre{\~n}o, S~Pratt, S.~Voloshin and N.~Xu. 
This research is supported by the U.S. Department of
Energy grant DOE-ER-40561. 
%

\end{document}